\documentclass[fleqn,usenatbib]{mnras}

\usepackage{newtxtext,newtxmath}
\usepackage[T1]{fontenc}
\usepackage{ae,aecompl}

\usepackage{hyperref}

\def\xlinkspace#1 #2{%
\ifx\relax#2%
\xlinkdash#1-\relax
\else
\xlinkdash#1 -\relax
\expandafter\xlinkspace\expandafter#2%
\fi}

\def\xlinkdash#1-#2{%
\ifx\relax#2%
\tmp{#1}%
\else
\tmp{#1-}%
\expandafter\xlinkdash\expandafter#2%
\fi}

\usepackage{graphicx}	
\usepackage{amsmath}	
\usepackage{amssymb}	
\usepackage{graphics,graphicx}
\usepackage{epstopdf}
\usepackage{color}
\usepackage{natbib}

\def\apj{\rm ApJ}
\def\apjl{\rm ApJL}
\def\apjs{\rm ApJS}
\def\aj{\rm AJ}
\def\mnras{\rm MNRAS}
\def\nat{\rm Nature}

\def\aap{\rm A\&A}
\newcommand{\msun}{M$_{\odot}$}

\title[Black Holes in Dwarf Galaxies]{Multimessenger Signatures of Massive Black Holes in Dwarf Galaxies}

\author[Bellovary et al.]{Jillian M. Bellovary$^{1,2}$\thanks{E-mail: jbellovary@amnh.org}, Colleen E. Cleary$^{2,3}$,Ferah Munshi$^{4,5}$,Michael Tremmel$^{6}$,  \newauthor Charlotte R. Christensen$^{7}$, Alyson Brooks$^{8}$, Thomas R. Quinn$^{9}$\\
$^{1}$Department of Physics, Queensborough Community College, City University of New York, 222-05 56th Ave, Bayside, NY, 11364\\
$^{2}$Department of Astrophysics, American Museum of Natural History, Central Park West at 79th Street, New York, NY 10024, USA\\
$^{3}$Department of Physics, Hunter College, City University of New York, New York, NY 10007, USA\\
$^{4}$Department of Physics and Astronomy, Vanderbilt University, 6301 Stevenson Center, Nashville, TN, 37212, USA\\
$^{5}$Department of Physics and Astronomy, University of Oklahoma, 440 W. Brooks St, Norman, OK, 73019, USA\\
$^{6}$Department of Astronomy, Yale University, 52 Hillhouse Ave, New Haven, CT, 06511, USA\\
$^{7}$Department of Physics, Grinnell College, 1115 8th Ave, Grinnel, IA, 50112, USA\\
$^{8}$Department of Physics \& Astronomy, Rutgers University, 136 Frelinghuysen Rd., Piscataway, NJ 08854, USA\\
$^{9}$Department of Astronomy, University of Washington, Box 351580, Seattle, WA 98122, USA
}
\date{Accepted XXX. Received YYY; in original form ZZZ}

\pubyear{2018}

\begin{document}
\label{firstpage}
\pagerange{\pageref{firstpage}--\pageref{lastpage}}
\maketitle

\begin{abstract}
Recent discoveries of massive black holes (MBHs) in dwarf galaxies suggest that they may have a more common presence than once thought.  Systematic searches are revealing more candidates, but this process could be accelerated by predictions from simulations.  We perform a study of several high-resolution, cosmological, zoom-in simulations focusing on dwarf galaxies that host massive black holes at $z = 0$, with the aim of determining when the black holes are most observable.  Larger dwarf galaxies are more likely to host MBHs than those of lower mass.  About 50\% of the MBHs in dwarfs are not centrally located, but rather are wandering within a few kpc of the galaxy center.  The accretion luminosities of MBHs in dwarfs are low throughout cosmic time, rendering them extremely difficult to detect.  However, the merger history of these MBHs is optimal for gravitational wave detection by LISA.  
\end{abstract}




\section{Introduction}

For years the existence of massive black holes (MBHs) in dwarf galaxies was thought of as an oddity, but recently several studies have uncovered evidence for a much more common presence \citep{Reines13,Moran14,Satyapal14,Lemons15,Sartori15,Pardo16}.  These discoveries pose several questions: what is the lower limit of galaxy mass which can host an MBH?  How and when do these MBHs form?  Which galaxies host MBHs and which do not, and why?  

The answers to these questions rest in the mechanism for forming MBHs, but this process is still shrouded in mystery.  Several MBH seed formation pathways have been proposed, including directly collapsing black holes in atomic cooling halos.  This scenario requires pristine halos with low angular momentum, such that the collapsing gas retains a large Jeans mass and does not fragment.   \citep{Loeb94,Eisenstein95,Haiman96,Oh02,Koushiappas04}.  In addition, the gas in this halo must remain atomic (with a temperature on the order of $10^4$K), because the formation of molecular hydrogen will enable further cooling (and thus fragmentation).  To prevent the formation of H$_2$, it is suggested that the halo may be irradiated by a nearby burst of star formation, producing UV light which dissociates any H$_2$ \citep{Dijkstra08,Shang10,Latif13a,Latif13b,Johnson14,Regan14,Choi15,Dunn18}.   The resulting collapsing object may first form a ''quasi-star'' \citep{Begelman06}, or possibly collapse directly into a black hole \citep{Lodato06}.  The resulting MBH seed is predicted to have a mass between $10^4 - 10^6$\msun~ and forms in halos with a virial temperature of $\sim 10^4$K \citep{Ferrara14}.

Other seed formation mechanisms may also be possible.  The remnants of Population III stars are expected to have masses of $\sim 100$\msun \citep{Couchman86,Madau01,Abel02,Heger02,Bromm04}, and may serve as building blocks for central supermassive black holes (SMBHs) through MBH-MBH mergers and gas accretion \citep{Micic07,Holley-Bockelmann10}.   Collapsing nuclear star clusters are also a promising avenue \citep{Begelman78,Devecchi09,Davies11,Katz15,Yajima16}.  Given the right circumstances of cluster metallicity, density, and dynamics, stars and/or compact objects within a cluster may collapse to form an MBH of $\sim 1000$\msun.  We wish to emphasize that more than one mechanism (or none of these!) may be at play, and the mechanism may vary with halo mass or other galaxy properties.  

One way to attempt to constrain the seed formation mechanism is through studying the occupation fraction of MBHs in the local universe.  While all galaxies with masses $M_{\rm gal} > 10^{10}$\msun~ seem to host MBHs, at lower masses the fraction seems to fall below one.  A true occupation fraction is impossible to measure, since inactive MBHs are only detectable through dynamical means in the nearest galaxies.  For more distant galaxies we must be content with an {\em active fraction}, i.e. those MBHs which are seen as active galactic nuclei (AGN).  Progress has been made in recent years, with the AMUSE survey constraining the active MBH fraction in early type galaxies with $M_{\rm star} < 10^{10}$\msun~ to be $\geq$ 20\%, and full occupation cannot be excluded \citep{Miller15}.  Extending such a study to late-type galaxies is more complicated, since the characteristically higher star formation rates will lead to a higher contamination in the X-rays from accreting binary systems.

Another way to constrain the seed formation mechanism is to study the smallest mass MBHs (currently $\sim 50,000$ \msun~ according to \citet{Baldassare15}), as well as the lowest mass galaxies which host MBHs.  These data can help us obtain an upper limit on the formation mass of MBHs, which can narrow the possibilities of formation mechanisms as we acquire larger samples.  Examining the mass, morphology, and stellar properties of dwarf galaxies with MBHs also gives us clues about the evolutionary history of the smallest MBH hosts, providing links to the formation history and co-evolution of MBHs and dwarfs.  Thus, finding and characterizing local dwarf galaxies hosting MBHs can shed light on this mysterious high-redshift process.

Dwarf galaxies are excellent laboratories for studying the interplay between MBHs and their hosts at the earliest times, when galaxies are small.  Critical to this subject is an understanding of the low-mass end of the MBH-galaxy scaling relations, such as $M-\sigma$.   The existence of the variety of tight scaling relations hints at coeval evolution between MBHs and their hosts \citep{Ferrarese00,Gebhardt00,Gultekin09,McConnell13}, although this relation may differ for active MBHs and quiescent ones \citep{Reines15}.  At the lowest masses there is very little data, but the data which exist hints at a larger scatter than for more massive hosts.  The placement of low-mass galaxies on diagrams such as $M-\sigma$ can help elucidate whether seeds are light or heavy \citep{VolonteriNatarajan} or whether MBH or stellar growth happens first.  

The current methods used to detect MBHs in dwarf galaxies rely on observational signatures from AGN 
\citep[see ][for a thorough review]{Reines16}.  Past works have shown success using narrow emission line ratios \citep{Reines13}, broad emission lines \citep{Reines13,Baldassare16}, nuclear variability \citep{Baldassare18}, and X-ray emission \citep{Reines11,Baldassare17} combined with radio emission \citep{Reines12}.    Each of these methods relies on there being detectable accretion onto the MBH.  MBHs which are not experiencing accretion events are difficult, if not impossible, to detect.  There are not likely to be enough stars within the sphere of influence to detect these MBHs dynamically, due to lower stellar surface densities in dwarfs combined with the lower MBH masses.  

In addition to electromagnetic signatures of MBH growth, the gravitational wave (GW) signal due to merging MBHs in small hosts may provide many clues for us about their existence and merger history.  While the occupation fraction of MBHs in dwarf galaxies is probably less than unity, dwarf galaxies are the most numerous type of galaxy in the universe.  
Thus, those that host MBHs and experience MBH-MBH mergers will contribute to the GW background signal.  The Laser Interferometer Space Antenna (LISA) will detect such mergers, and help us characterize the number and masses of MBHs which form in small galaxies.   
LISA is optimized to detect merging MBHs with a range of total masses from $10^4 - 10^7$\msun~\citep{LISA}, which is specifically the range one expects to find MBHs in dwarf galaxies.  GW signal predictions have been made from the Illustris simulation \citep{Kelley17}, however the majority of the relevant SMBHs in Illustris are larger than LISA's detectable range, and are instead a good fit for pulsar timing array measurements. 

There is little prior work on simulations of dwarf galaxies hosting MBHs.  In \citet{Bellovary10} the authors discussed how dwarfs hosting MBHs may be tidally disrupted when merging with larger disk galaxies, depositing their MBHs in the larger galaxy halo; however, no study of the pre-merger dwarfs was done.  More recently, \citet{Habouzit17} simulated a volume of low-mass galaxies and low-mass seed MBHs, demonstrating their likely evolution onto the $M-\sigma$ relation; however, these simulations stop at $z = 3$.
Thus, high-resolution simulations of dwarf galaxies hosting MBHs at $z = 0$ are an important step in furthering our understanding of how MBHs interact with their dwarf galaxy hosts.

In this paper we present the first high resolution sample of simulated dwarf galaxies hosting massive black holes.  Our sample consists of field dwarfs as well as satellites of Milky Way-type galaxies.  A full analysis of the simulated field dwarfs will be done by Munshi et al. (in prep), while the satellite sample is being presented for the first time here.  We explore the observability of the dwarf-hosted MBHs with electromagnetic signatures due to accretion, and predict the events LISA will detect due to GW radiation.

\section{Simulations}

Our simulations were run with the state-of-the-art N-body Tree+SPH code \textsc{ChaNGa} \citep{Menon15}.  \textsc{ChaNGa} incorporates the Charm++ framework resulting in improved scalability, with the ability to run on 100,000+ cores.  Our simulations retain the same physics modules as \textsc{ChaNGa}'s precursor, \textsc{Gasoline} \citep{Stadel01,Wadsley04}.  These include a cosmic UV background \citep{Haardt12}, star formation based on molecular gas fraction \citep{Christensen12}, blastwave supernova feedback \citep{Stinson06}, and metal diffusion and cooling \citep{Shen10}.  Our simulations also incorporate the MBH formation, growth, and feedback models described in \citet{Tremmel17} as well as a prescription for MBH dynamical friction \citep{Tremmel15}.  We use a modernized calculation of the SPH kernel by using a geometric mean density in the SPH force expression \citep{Ritchie01,Menon15,Wadsley17}, which accurately reproduces shearing flows such as Kelvin-Helmholtz instabilities.  Our prior work using these models has been very successful in reproducing detailed characteristics of galaxies which match observed scaling relations, including the stellar mass - halo mass relation \citep{Munshi13}, the satellite galaxy distribution of massive galaxies \citep{Zolotov12,Brooks14}, the Kennicutt-Schmitt relation \citep{Christensen12}, and the mass-metallicity relation \mbox{\citep{Brooks07,Christensen16}}.

\subsection{Simulation Properties}

We present two categories of dwarf galaxy populations, which are selected from two sets of initial conditions.  Both sets of simulations use the ``zoom-in'' volume renormalization technique of \citet{Katz93} to resimulate pre-selected galaxies from a uniform volume at high resolution.  Our sample of field dwarfs from low-density environments, known as the ``MARVEL-ous Dwarfs,'' consists of 64 galaxies, and originates from a 25 Mpc volume, which uses cosmological parameters from WMAP3 \citep{WMAP3}.   These simulations have a force softening resolution of 60 pc, dark matter particle masses are 6660\msun, while gas particles have a mass of 1410\msun~ and star particles 422\msun.  The full sample of the MARVEL-ous dwarfs will be fully introduced and described in an upcoming paper (Munshi et al. in prep).    The other sample of dwarfs, which consists of 101 galaxies, are in environments near Milky Way-like disk galaxies, and are selected from a 50 Mpc volume using Planck cosmological parameters \citep{Planck}.  The Milky Way halos were selected to mimic the Milky Way in terms of mass and morphology, but with a variety of assembly histories, and are named the ``DC Justice League.''  These simulations have a force softening resolution of 170 pc, dark matter particle masses are $4.2 \times 10^4$\msun, while gas particles have a mass of $2.7 \times 10^4$\msun~ and star particles 8000 \msun.  

In both cases, we identify halos using the Amiga Halo Finder \citep{Knollmann09} which uses an overdensity criterion for a flat universe \citep{Gill04}.  In this study, we only consider halos that we define as resolved. A resolved halo has at least 1500 dark matter particles and an extended star formation event lasting at least 100 Myr. In such an event, a halo forms a burst of stars, those stars pollute the interstellar medium, and the halo forms a subsequent burst of stars.  We do this because: (1) extended star formation histories are typical of observed dwarf spheroidal galaxies, and (2) this criterion is more physical than making a cut in number of stars. In particular, by doing so, we are ensuring that we are not simply picking up stochasticity or noise from our subgrid star formation models, especially in halos which are in the ultrafaint regime and near our resolution limits. Note that neither of these criteria are restrictive in the range of stellar or halo masses which describe MBH hosts.    

While the cosmological parameters for the two sets of simulations differ, we do not expect substantial differences between our two sets of galaxies as a result.  The cosmological evolution of dwarfs, MBH formation, and  star formation histories are far more dependent on environment than cosmological parameters, and will exhibit no measurable difference between the two cosmologies.  The simulations also differ in resolution; however the masses at which the MBH particles form are the same.   The primary concern in the low resolution runs is the accuracy of our dynamical friction prescription; however,  \citet{Tremmel15} shows that at this resolution the dynamical friction prescription has converged.  We discuss the repercussions further in Section \ref{sect:BHphysics}.

Both sets of simulations use the same physical modules and parameters for star formation and supernova feedback.  Stars are formed using the recipe described in \citet{Christensen12}, which weights the star formation probability by the molecular hydrogen fraction.  Stars are allowed to form if the gas particle density $\rho > 0.1$  cm$^{-3}$, though the actual density of star-forming gas is commonly 100 - 1000 cm$^{-3}$ due to the required fraction of H$_2$.  Additionally, temperature must be $T < 1000$ K, and we set the star formation efficiency parameter $c^*_0 = 0.1$.  Each star particle is represented with a Kroupa initial mass function (IMF) \citep{Kroupa}.  We use the Blastwave supernova recipe described in \citet{Stinson06}, which disables cooling for the theoretical lifetime of the momentum conserving phase, with a supernova energy value of $E_{\rm SN} = 1.5 \times 10^{51}$ erg.

\subsection{Black Hole Physics}\label{sect:BHphysics}
All simulations presented here use the MBH physics modules presented in \citet{Tremmel17}, and we summarize the main details here.  Black hole particles form in extremely overdense regions 
(
e.g. 3000 cm$^{-3}$ for the lower-resolution simulations and $1.5 \times 10^4$ cm$^{-3}$ for the higher-resolution runs).  Additionally, they must exhibit low metallicity ($Z < 10^{-4}$), low molecular hydrogen fraction ($f_{\rm H_2}< 10^{-4}$), and a maximum temperature of $2 \times10^4$K.  The gas particle must also exceed a Jeans Mass criterion, i.e. $M_{\rm Jeans} =  (\pi^{5/2}c^2) / (6\rho^{1/2}) > 4 x 10^5$\msun~ in the lower-resolution case and $> 10^5$\msun~ in the high-resolution case.  This criterion ensures that the particle is in a region which is likely to collapse.

These criteria rely on local gas properties only, and do not depend on the gas ``knowing'' about global halo properties or halo occupation fractions.  We can thus broadly reproduce the criteria for direct collapse black hole seed formation, ensuring that the gas in question is collapsing quickly while cooling relatively slowly \citep{Begelman06,Volonteri12}.  Once formed, the MBH particle accretes mass from the surrounding gas, until it either depletes its neighbourhood of gas or reaches a mass of 50,000 \msun, whichever happens first.  This process represents a period of rapid, unresolved black hole growth.  Since the two sets of simulations have different resolutions, in practice the seed masses differ by about a factor of two.  The Justice League (satellite) simulations, with lower resolution, easily reach the 50,000 \msun~ value by accreting a small number of nearby gas particles.  The MARVEL-ous dwarfs, on the other hand, are at higher resolution and the nearby gas particles are smaller as a result.  Most MBH particles in these field dwarf simulations reach around 25,000 \msun~ before they run out of neighbours to accrete.  This factor of two difference is of little significance, since the true mass of direct collapse black holes is unknown to within a few orders of magnitude.  In comparing one set of simulations to another, the MBHs interact with their hosts and each other in the same way, so there is little quantitative difference due to this mass disparity.

We do not attempt to resolve true MBH mergers, only close MBH pair formation.  When a close pair reaches a separation that is (typically) well below our resolution limit, it is appropriate to treat the pair as a single particle.  Black hole particles merge if they are close in space (within two softening lengths) and have low relative velocities.  Specifically, they must meet the criterion $\frac{1}{2}\Delta {\vec{ \rm v}} < \Delta {\vec{ \rm a}} \cdot \Delta {\vec{ \rm r}}$, where $\Delta {\vec{ \rm v}},  \Delta {\vec{ \rm a}}$ and  $\Delta {\vec{ \rm r}}$ represent the relative velocity, acceleration, and radius vectors of the two MBHs respectively.  This prescription mimics the unresolved condition of the two MBHs being gravitationally bound to each other.  The binary hardening timescale for MBH binaries is uncertain, but likely to be on the order of $10^7 - 10^8$ years \citep{Armitage02,Haiman09,Colpi14,Holley-Bockelmann15}.  These timescales are small compared to the relevant timescales of the simulation, and so it is appropriate to treat the close pair as a single merged MBH upon meeting the relevant criteria.

We implement the sub-grid dynamical friction (DF) model for MBHs described in \citet{Tremmel15}, which gives an estimate for DF on scales smaller than the gravitational softening length based on the Chandrasekhar formula \citep{Chandrasekhar43,Binney08}.  This formalism allows us to accurately follow the dynamics of MBHs rather than artificially pinning them to the centers of their host galaxies.  As a result, we can realistically track the orbital evolution of MBHs during galaxy mergers and the subsequent MBH pairing and merger (see \citet{Tremmel17b} for a thorough study of MBH pairing in the \textsc{Romulus} simulation).    

This sub-grid model is effective in allowing MBH orbits to decay with a realistic dynamical time, as long as the MBH particle has a mass that is $\sim 3$ times larger than the surrounding particles.  While the minimum mass in the lower-resolution simulations does not always fulfill this criterion, mergers with other MBHs cause most of the black holes to grow above this limit.  We specify in \S \ref{sect:results} the few instances in which our results may be affected by unresolved dynamical friction.

Black holes grow by accreting gas from the surrounding environment.  Thermal energy from accretion is distributed isotropically along the SPH kernel to the nearest 32 particles.  Gas particles which receive feedback energy are not allowed to cool for the duration of their own timestep (usually $10^3 - 10^4$ years), which mimics the continuous deposition of feedback energy.  The feedback energy is directly proportional to the accretion rate, and we assume a radiative efficiency $\epsilon_r$ = 0.1 and a feedback coupling efficiency of $\epsilon_f$ = 0.02.  The latter parameter was determined using a multi-parameter optimization technique (see \citet{Tremmel17} for an in-depth discussion).  The parameter optimization was done at a lower resolution than the simulations we present here.  Because the MBHs accrete very little (see Section \ref{sect:results}), however, we do not expect that free parameters related to accretion and feedback to have a strong effect on our results.

We model accretion using a modified version of the Bondi-Hoyle formula.  In calculating the accretion rate $\dot M$, we consider the nearest 32 gas particles.  The Bondi-Hoyle formula assumes a spherically symmetric, non-rotating flow, which is an unlikely scenario in an actual galaxy.  Realistically, gas which is inflowing radially may be efficiently accreted, but gas with a substantial tangential velocity component has too much angular momentum to fall into the MBH.  To address this issue, we have developed a modification to the Bondi-Hoyle formula that accounts for the angular momentum of gas on resolved scales.  We include an additional term for rotational velocity, $v_{\theta}$, in the Bondi-Hoyle equation to account for the rotational component of the gas, which effectively reduces the accretion rate.  This term is used in place of the regular bulk velocity term $v_{\rm bulk}$ as long as $v_{\theta} > v_{\rm bulk}$ (otherwise $v_{\rm bulk}$ is used).  Additionally, we use the boost factor first described in \citet{Booth09}, which includes a density-dependent multiplier to the accretion rate, depending on whether the situation is well-resolved.  We use a boost factor of $\beta = 2$ in the case where the local gas density is high enough to be unresolved.  In summary, the accretion rate is calculated as follows:
 
 \begin{equation}
 \dot M =\left\{
  \begin{array}{@{}ll@{}}
 \frac{\pi G^2 \alpha M_{\rm BH}^2 \rho}{(v_{\rm bulk}^2 + c_s^2)^{3/2}}, & \text{for $v_{\rm bulk} > v_{\theta}$} \\
 \frac{\pi G^2 \alpha M_{\rm BH}^2 \rho c_s}{(v_{\theta}^2 + c_s^2)^2}, & \text{for $v_{\rm bulk} < v_{\theta}$} \\
   \end{array}\right.
 \end{equation}
 
 \noindent
 where $\alpha$ = 1 if the local gas density is less than the star formation threshold density $\rho_{\rm th}$, and $\alpha = (\rho / \rho_{\rm th})^{\beta}$ in the instance when it is greater.
 
This combination of MBH subgrid models has been very successful in reproducing a variety of observed scaling relations and properties.  The \textsc{Romulus} simulation is a 25 Mpc volume which demonstrates results consistent with the $M_{\rm BH}-M_{\star}$ relation, the stellar mass - halo mass relation, as well as cosmic star formation and MBH accretion histories \citep{Tremmel17}.  Our work here is the first examination of this prescription in high resolution zoom-in simulations.
 
 \subsection{Limitations of the Model}\label{sect:Limitations}
 
\begin{figure}
\includegraphics[width=\columnwidth]{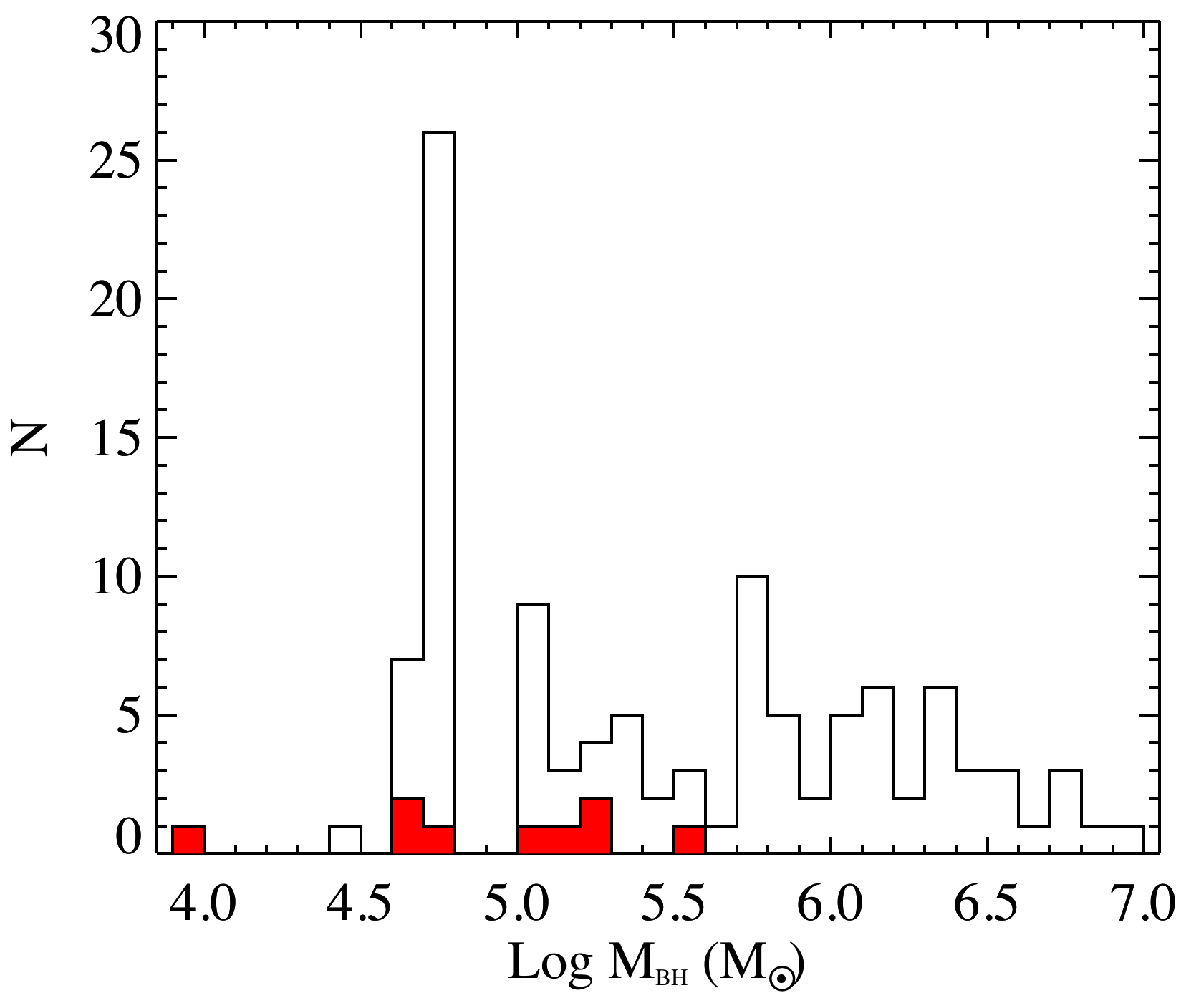}
\caption{ The black histogram shows the distribution of the initial MBH masses in all simulations.  The red filled histogram represents the subset of MARVEL-ous dwarf MBHs.  There is a large peak at $\sim 50,000$\msun, which is the seed mass for both simulations.  Higher values are due to immediate post-formation MBH mergers (see Section \ref{sect:Limitations}).  \label{fig:IMF}
}
\end{figure}

We set the minimum mass of a MBH particle at the time of seeding, which is accomplished by the rapid consumption of nearby gas particles.  Because seeding occurs in the densest regions, multiple MBH particles may form in the same locality at the same time.  As a result they often coalesce rapidly, resulting in one MBH particle with the total mass of the progenitors, which gives an effective IMF for the MBHs.  These mergers are not true mergers, but reflect the rapid growth of a single MBH in a dense environment, which is thought to be a rapid process and below our resolution limit \citep{Hosokawa13,Schleicher13}.  In this analysis, we define a spurious merger as one which occurs within 100 Myr of the formation of either MBH in the pair.   This time frame is long enough for feedback due to MBH accretion to heat the gas surrounding an MBH, preventing the gas from fulfilling the temperature criteria for seed formation.  Any MBH which survives this time demarcation is subsequently considered for further analysis in our study.  We show the distribution of the MBH masses {\em after} spurious mergers in Figure \ref{fig:IMF}.  The black histogram is the total distribution of masses, while the red filled histogram is those in the MARVEL-ous dwarfs sample only.
 
Due to the uncertainty in initial masses, in this paper we focus on quantities that are not strongly dependent on precise MBH mass values, but rather on the existence of MBHs in dwarf galaxies in general.   All MBH masses (and subsequent estimated luminosities) can be taken as upper limits.  The uncertainty in masses will affect our specific predictions for LISA (see Figure \ref{fig:waterfall}, Section \ref{sect:detect}), but otherwise have little relevance to our results.

Recent simulations have shown that small changes in modeling supernova feedback can have substantial results on galaxy and/or MBH evolution.  For example, \citep{Habouzit17} examine how three different feedback models affect the MBH occupation fractions and growth in a set of cosmological simulations, and find different results for varying models.  In general, varying feedback models can change outflow rates and densities, altering the evolution of a galaxy \citep[e.g.]{Rosdahl17}.  Our subgrid models, including blastwave feedback, have been shown to produce realistic galaxies which obey observed relations such as Tully-Fisher \citep{Governato09}, stellar mass - halo mass \citep{Munshi13}, mass-metallicity \citep{Brooks07,Christensen14}, and produce dwarf and Milky Way-size galaxies with realistic morphologies and formation histories \citep{Christensen14,Christensen16,Brooks17}.  We are confident that our results strongly represent the reality of the physical universe.  In terms of MBHs, supernova feedback will have little to no effect on their formation, since gas which forms MBHs must have essentially zero metallicity.  Thus all MBHs form before local supernova occur.  The question of whether supernova feedback limits MBH growth is an open one, since our MBHs do not grow rapidly.  However, since MBHs in the Romulus simulation (with the same feedback recipe) do grow efficiently, even in low mass galaxies, supernovae to not seem to expressly prohibit MBH growth.  More exploration is needed on this topic before we settle upon a firm answer.

\section{Properties of MBHs and their Dwarf Galaxy Hosts}\label{sect:results}

\subsection{MBH Properties}\label{sect:properties}

\begin{figure}
\includegraphics[width=\columnwidth]{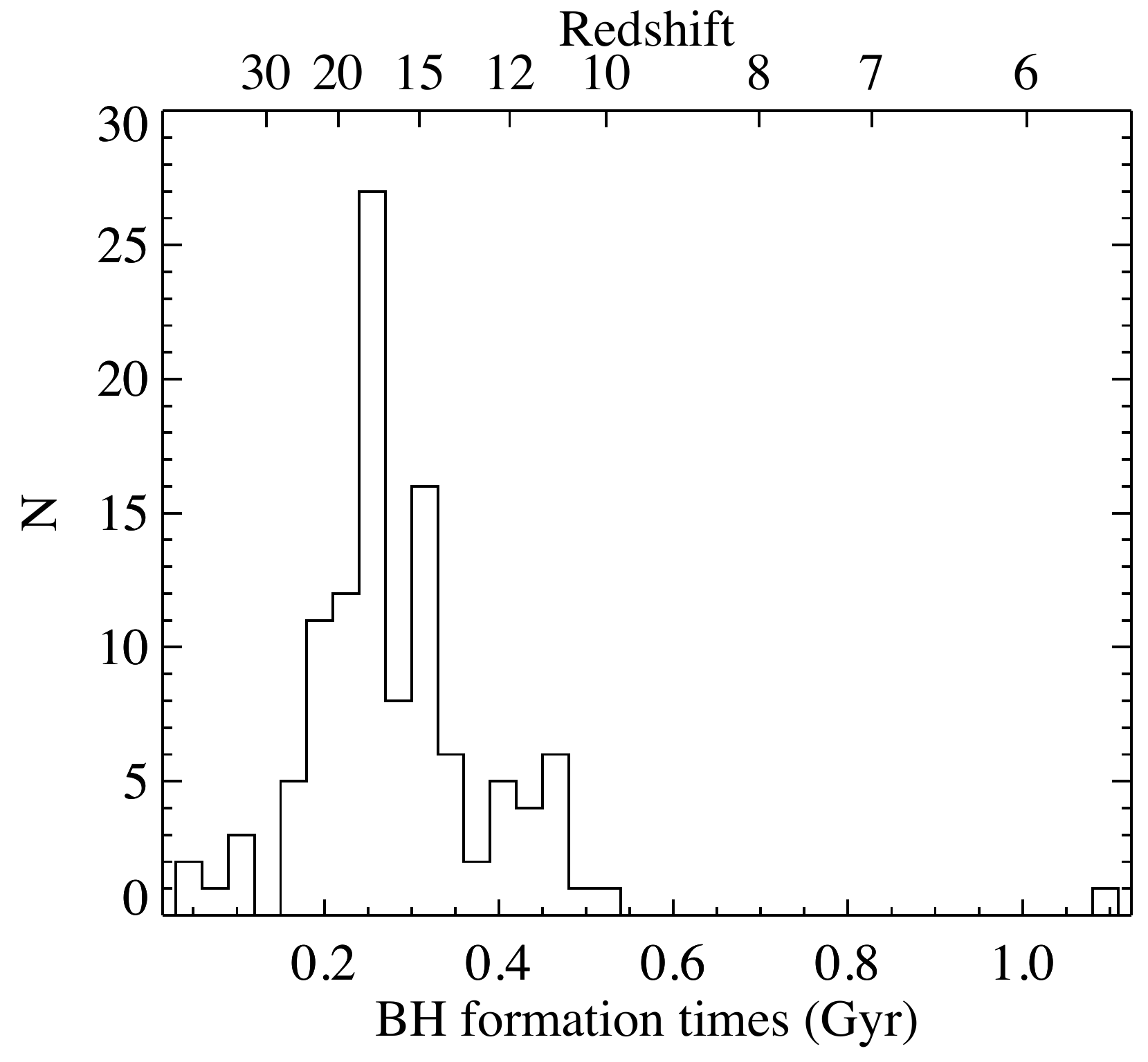}
\caption{ Distribution of the formation times of all MBHs in all combined simulations.  MBHs form in a burst in the first few Myr years of the universe, and then formation drops off sharply.  Our results are consistent with theories of direct collapse black hole formation, which predict formation redshifts from $6 < z < 20$ \citep[e.g.][]{Begelman06}.  \label{fig:formtimes}
}
\end{figure}

MBH seeds form at early times in low-mass halos.  Formation can begin as early as $z \sim 50$, and peaks in the range of $10 < z < 20$.This result is consistent with our prior work using a less sophisticated model \citep{Bellovary11}.  In Figure \ref{fig:formtimes} we show the distribution of times that the MBHs form for all simulations combined.  
The truncation of seed formation is a result of the propagation of metals in the interstellar medium due to supernova feedback.  This metallicity criterion ensures that MBH seed formation is a high redshift phenomenon; any MBH that forms at lower redshift must exist in a pristine pocket of the universe; while such a thing is possible, our simulations in this work do not exhibit this behavior.

\begin{figure}
\includegraphics[width=\columnwidth]{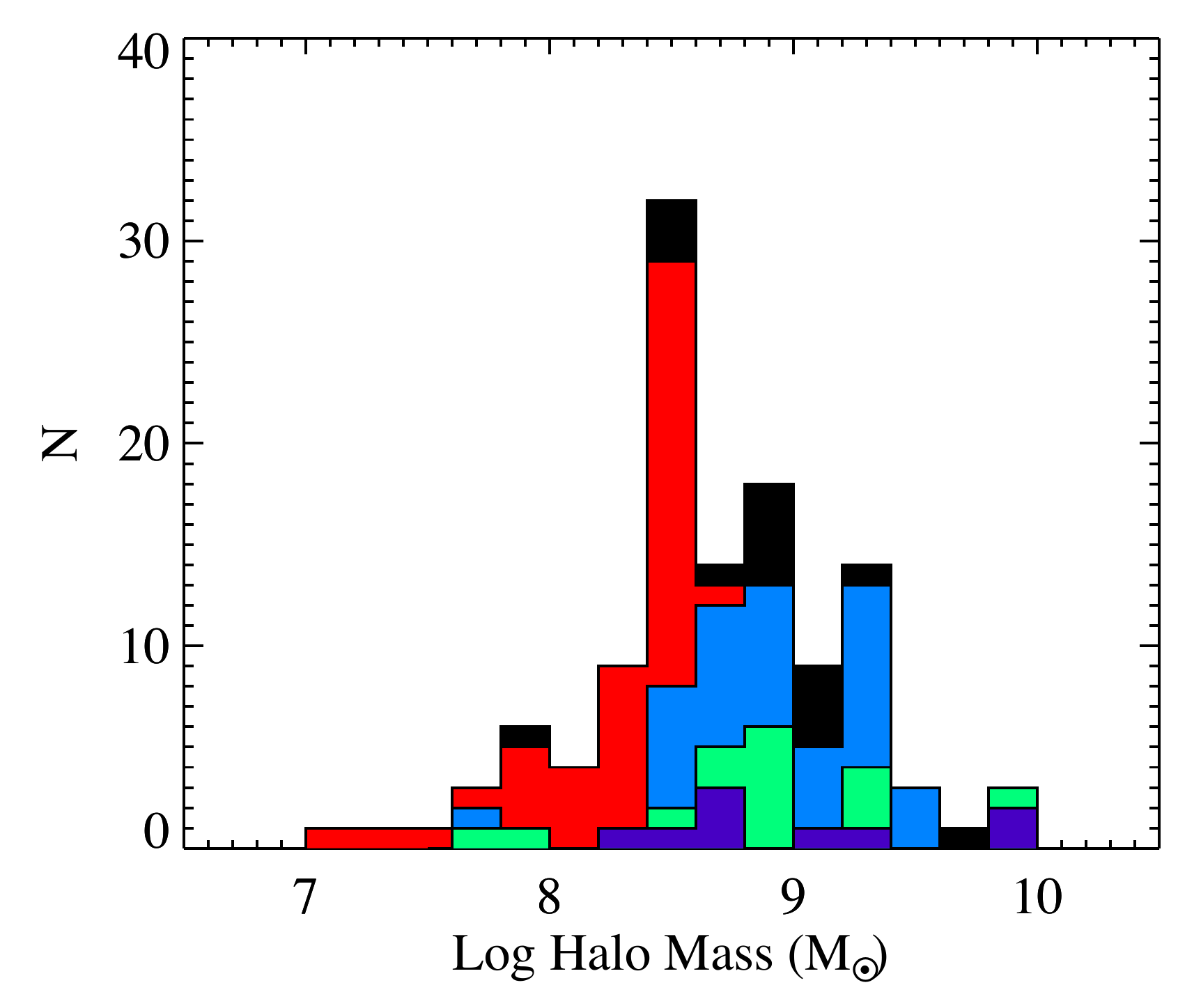}
\caption{Distribution of the log of halo mass at the time the MBH formed.  Each colour represents a distribution from one simulation (and the dark purple is all of the MARVEL-ous dwarfs together), and are stacked to represent the total distribution.
Halo masses at formation are mainly in the $ 8 < $ log$M_{\rm vir} < 9.5$ range.  Direct collapse black hole formation is expected to take place in halos with masses of $\sim 10^7 - 10^9$\msun~ \citep[e.g.][]{Begelman06}.  \label{fig:formmass}
}
\end{figure}

MBHs form in halos primarily with masses in the range of $ 10^8 < M_{\rm vir} < 10^{9.5}$~\msun.  Figure \ref{fig:formmass} shows the virial mass distribution of each host halo at the time the MBH forms.  The complete histogram represents the sum of the results of each simulation, and peaks at a value of log($M_{\rm vir}) \sim 8.5$.  The coloured portions of the histogram represent each individual simulation; each of these echoes the broader distribution.  This halo mass range is expected from models of direct collapse; the temperature at which virialized halo gas reaches a temperature of $10^4$K occurs within this mass range at redshifts of $10 < z < 20$ \citep{Begelman06,Lodato06}.  We emphasize that our MBH formation model is independent of global halo properties; MBHs do not form in halos of a particular mass by design.  Rather, our model based on local gas properties is an appropriate representation of the physics processes thought to govern MBH formation.  This model allows for MBHs to form in relatively low-mass halos, which is critical for our study of MBHs in dwarf galaxies.

\begin{figure*}
\begin{center}
\includegraphics[scale=0.5]{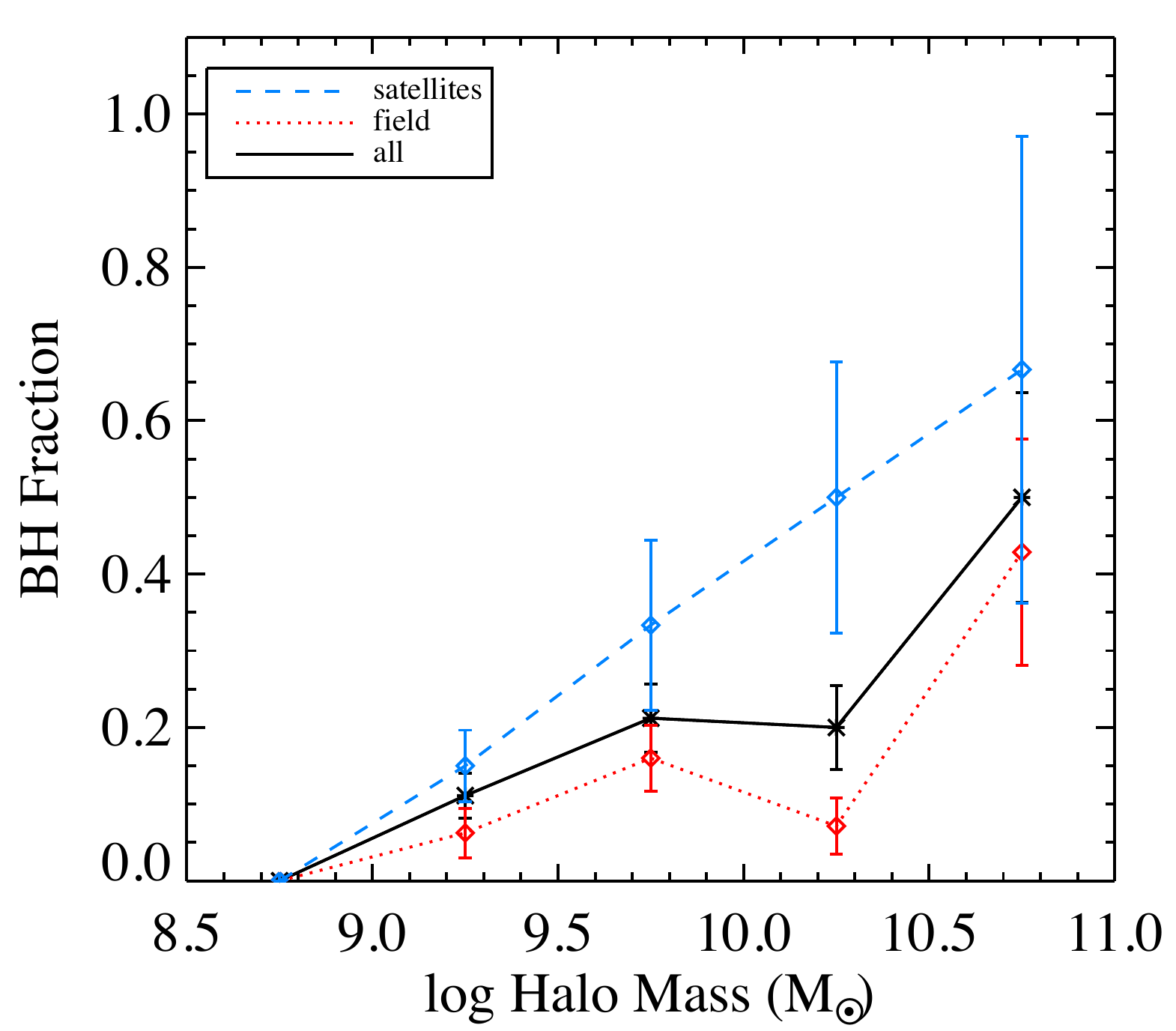}
\includegraphics[scale=0.5]{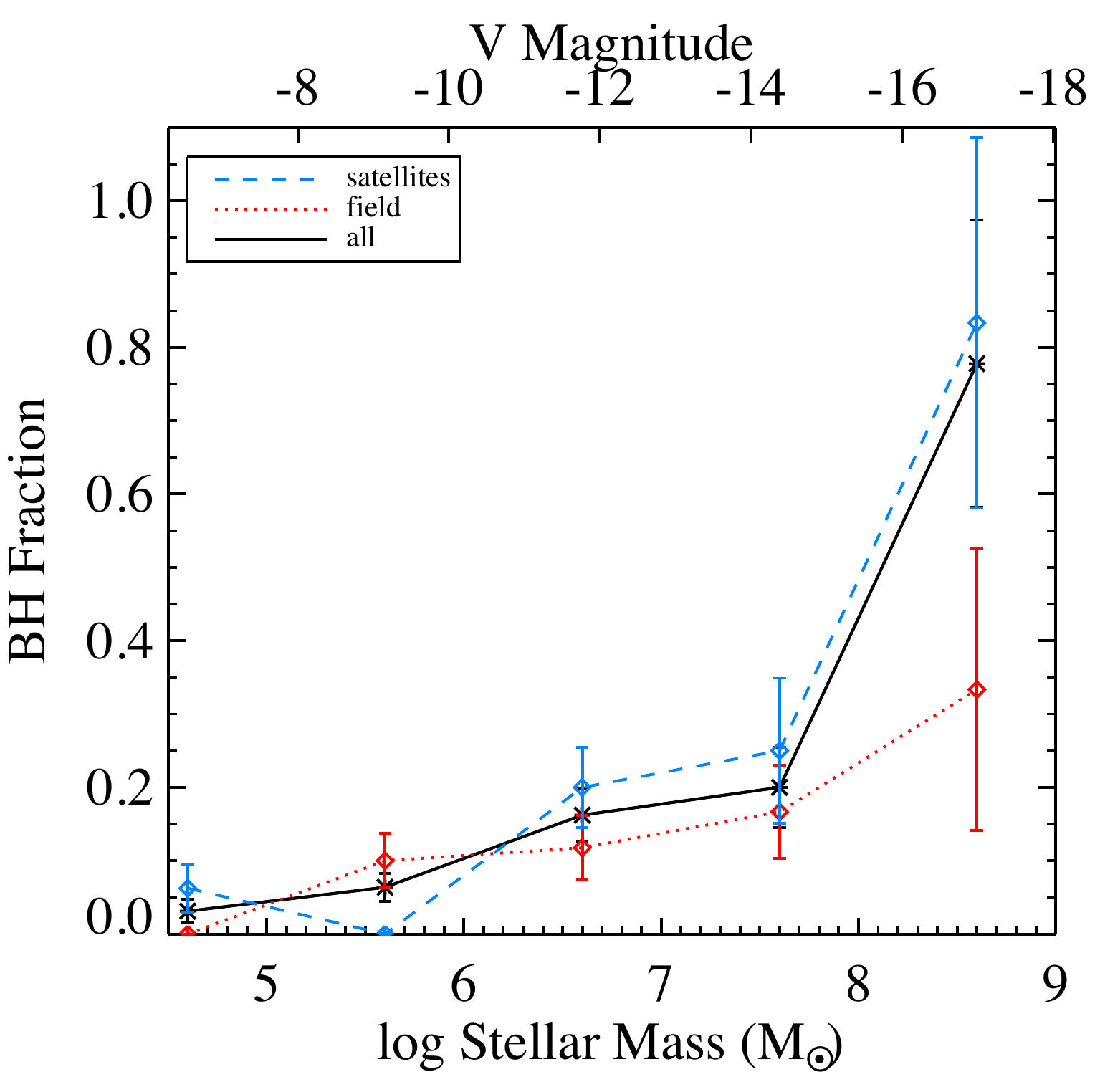}

\caption{ Occupation fraction of MBHs.  The left panel is vs. virial (total) mass, where the black solid line represents all halos which meet our resolution threshold.  The blue dashed and red dotted lines represent satellite and field populations, respectively.  Error bars are Poisson errors.  The right panel is vs. stellar mass, and the top axis is the corresponding absolute $V$ magnitude (see text for details).    
  \label{fig:occfrac}
}
\end{center}
\end{figure*}

MBHs do not exist in every dwarf galaxy, but they are more likely to be hosted in more massive dwarfs.
Figure \ref{fig:occfrac} shows the $z = 0$ occupation fraction of MBHs in terms of both virial mass (left) and stellar mass (right).  Our results are consistent with \citet{Miller15}, who constrain the active fraction of early-type galaxies with stellar masses less than $10^{10}$\msun~ to be greater than 20\% at $z = 0$.  For both figures, the black line represents the total occupation fraction, while the red and blue lines divide the sample into field and satellite galaxies, respectively.    A satellite is defined as being within a distance of $2R_{\rm vir}$ from a large (Milky Way-size) halo at $z = 0$.  Error bars represent Poisson errors.   In our simulations, the formation of MBH particles is inherently stochastic due to the nature of our star formation recipe and the likelihood that gas will reach the large overdensity required for MBH candidacy.  As a result, not all galaxies will host MBHs, but there is a higher occupation fraction for galaxies which are more massive.  

The right panel of Figure \ref{fig:occfrac} shows the occupation fraction for a given stellar mass, which shows a significant increase in the fraction of MBH hosts between $10^8$ and $10^9$\msun.  For this figure, the bottom axis represents the stellar mass directly from the simulation multiplied by a factor of 0.6, which corrects for observational limitations in observing low surface-brightness features \citep{Munshi13}.  The top axis is the inferred absolute $V$ band magnitude. 
Based on the age and metallicity of each star particle, we used the Starburst99 stellar population synthesis models of \citet{Leitherer99} and \citet{Vazquez05}, adopting a \citet{Kroupa} IMF, and calculated the $V$ magnitudes and $B - V$ colours of each galaxy.  Dust reddening was not taken into account; however dwarf galaxies are rarely dust-rich and the effect of redding should be small.  We verified that these magnitudes return a similar stellar mass to that of our simulation by recalculating the stellar mass using the mass-to-light-ratio formula from \citet{Bell00}, and inputting the colours and luminosity from our calculation.  Confident in our luminosities, we calculated a linear fit for values of $V$ for a given simulated stellar mass to acquire the axis label at the top of Figure \ref{fig:occfrac} (right panel).  
   
There is an apparent environmental difference in MBH-hosting dwarfs when looking at total halo mass, but this difference seems to disappear with stellar mass.  (The difference persists at the highest mass bin, but the error bars are large due to there being very few objects in this bin.)  This difference can be explained by the tidal stripping of satellites as they enter larger halos.  The satellites of Milky Way-like galaxies, which exist within 2 $R_{\rm vir}$ of their hosts, have experienced tidal stripping, which preferentially removes dark matter from the outskirts of each halo.  This process decreases the virial (halo) mass, while keeping the stellar mass roughly intact.  As a result, the MBH occupation fraction appears to shift upwards for satellites of lower halo masses.  In reality, these halo masses have been lowered due to environmental effects, and so the distribution has actually shifted to the left.  MBHs do not preferentially form in halos near larger overdensities; rather, they form independently of large-scale environmental density.

\begin{figure}
\includegraphics[width=\columnwidth]{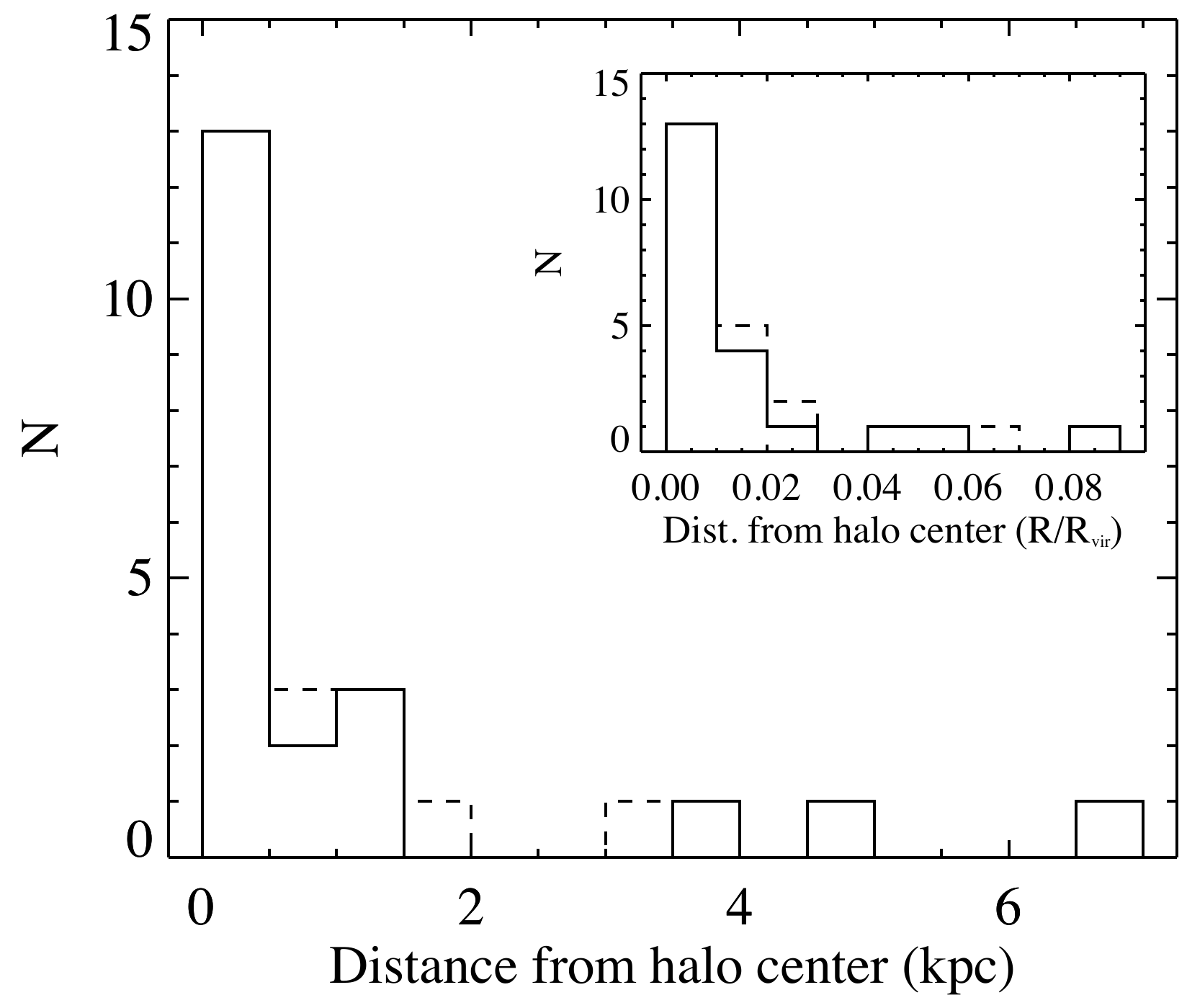}
\caption{The radial distribution of MBHs in their hosts.  The main figure shows distance in kpc, and the inset in terms of $R/R_{\rm vir}$.  About half of the MBHs are not centrally located.  The solid line represents all MBHs for which we are confident dynamical friction is properly resolved; the dashed line includes three additional objects which may be off-center due to resolution effects.  \label{fig:centers}
}
\end{figure}

Turning our discussion to the $z = 0$ properties of the galaxies and MBHs, we first present the locations of the MBHs in each dwarf galaxy (Figure \ref{fig:centers}), in terms of both physical distance and scaled by virial radius (inset).  A substantial fraction of MBHs in dwarf galaxies are not located in the galactic centers; rather, the population is divided between central and ``wandering'' MBHs.    Dwarf galaxies lack a deep potential well, which would stabilize the MBH's position in the center; these dwarfs tend to exhibit cored density profiles, and thus the shallow potential well gives MBHs more ``room'' to exist within \citep{DiCintio17}.   Dwarf galaxies often have irregular morphologies as well, and defining the center is not always straightforward.  We define each galaxy's center using the concentric shrinking sphere method described in \citet{Power03}.  

Our subgrid model for dynamical friction (see Section \ref{sect:BHphysics}) ensures that the dynamical evolution of MBHs are tracked self-consistently and accurately; thus, MBHs which are located in galaxy centers are realistically expected to be there, while those located in galaxy outskirts are there due to their unique dynamical histories.  There are three off-center MBHs which have less than the maximum mass for resolved dynamical friction, however, and we are not confident that their positions are due to physical effects (see \S \ref{sect:BHphysics}).  We have depicted the full distribution of MBHs with a dashed line in Figure \ref{fig:centers}, and the distribution without these objects as the solid line.

Dynamical perturbations are common, causing the MBHs to vacate the galaxy center.  Once the MBH has left the galaxy center, dynamical friction is often not efficient enough to bring it back, since the background density of stars and dark matter is relatively low.  Dynamical friction would be efficient in a more massive galaxy like the Milky Way, with a much higher stellar density, but for the dwarfs the time for the orbital decay is often greater than the Hubble time.  To confirm this assumption, we estimate the dynamical friction timescale for our wandering MBHs using the formula derived in \citet{Binney08}:

\begin{equation}
t_{\rm df} \sim \left(\frac{19~  {\rm Gyr}}{{\rm ln} \Lambda}\right)\left(\frac{r_i}{5~ {\rm kpc}}\right)\left(\frac{\sigma}{200~ {\rm km/s}}\right)\left(\frac{10^8 {\rm M}_{\odot}}{ M_{\rm BH}}\right)
\end{equation}

We assume $\Lambda \sim b_{\rm max}/b_{\rm min}$ with the maximum impact parameter, $b_{\rm max}$, equal to the $z = 0$ radius of the MBH orbit, $r_i$.  We directly measure the velocity dispersion, $\sigma$, of all of the particles within $r_i$ in each halo.  We use a minimum impact parameter, $b_{\rm min}$, of 10 pc to represent the characteristic size of nuclear star clusters, which commonly exist around MBHs in this mass range.  We also add a factor of 50\% to the mass of each MBH, as an estimate for the additional mass of this putative cluster.  We define non-central MBHs as being more than 400 pc from the galaxy center (just over two softening lengths for the low-resolution simulations).  70\% of the non-central MBHs in dwarfs have dynamical friction timescales longer than 10 Gyr, confirming our assumption that once they leave the center they do not return.

It is worth pointing out that some of the best intermediate mass black hole candidates are off-nuclear ultraluminous xray sources.  Objects such as HLX-1 \citep{Farrell09}, NGC 2276-3c \citep{Mezcua15}, and the ULX in NGC 5252 \citep{Kim15} are all consistent with dwarf galaxy nuclei stripped during a merger.  Mergers may be a common way to perturb MBHs from galaxy centers, though the MBHs may not always be accreting during this process.

\subsection{Host Galaxy Properties}

We investigate whether dwarfs which host MBHs have noticeably different properties compared to those which do not.  Feedback from an accreting MBH may have a substantial impact on the gas distribution within a galaxy, which may in turn affect the quenching of star formation \citep{Bradford18}.  

\begin{figure}
\includegraphics[width=\columnwidth]{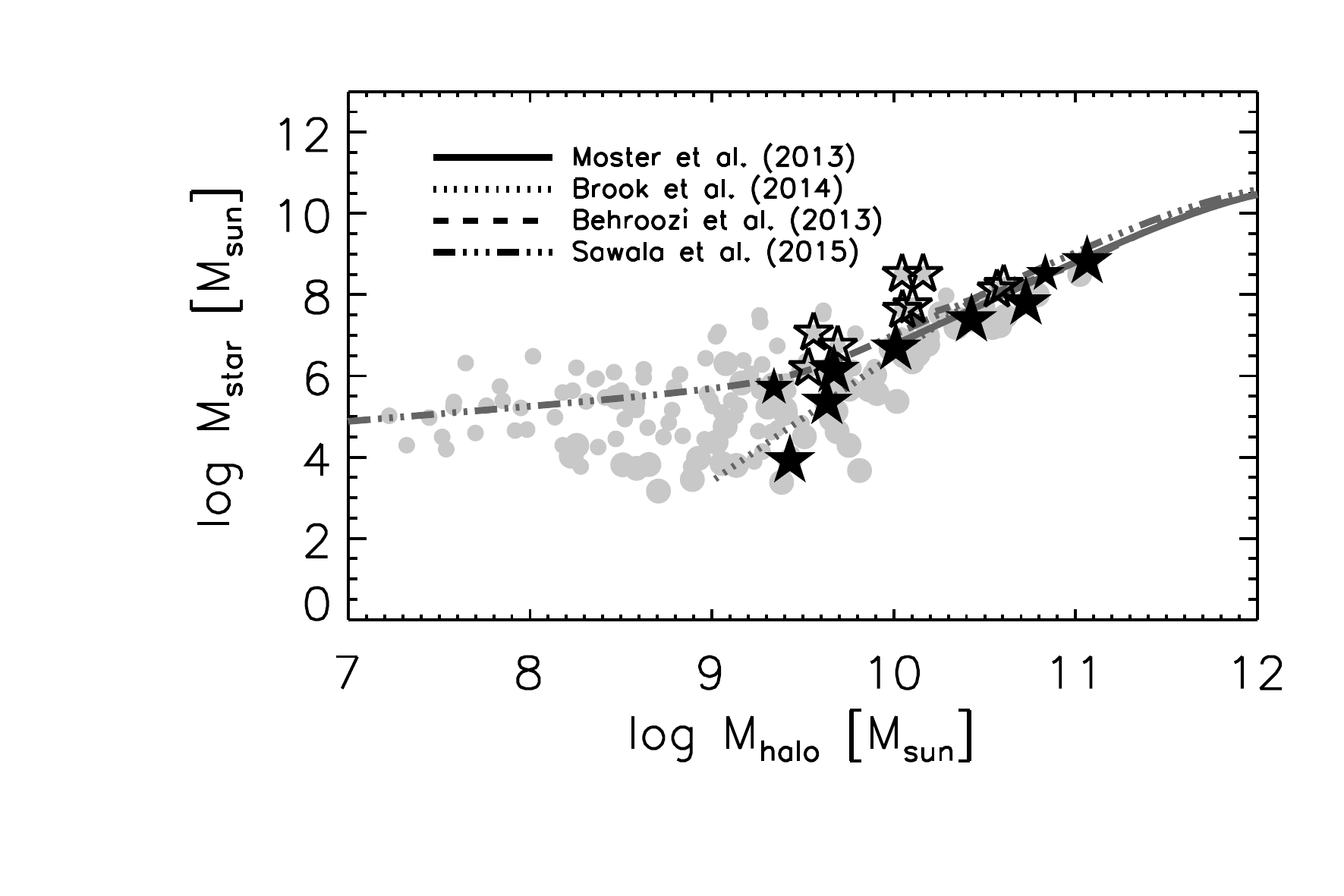}
\caption{ Stellar to halo mass of the full sample of dwarfs, highlighting the black hole dwarf hosts.  All halos are shown greyed out in the background in circles and dwarfs hosting black holes are shown with black stars.  Filled stars correspond to isolated or ``field'' dwarfs, and open stars are satellites.  Symbol sizes correspond to mass resolution -- smaller symbols are the lower resolution Justice League dwarfs, while larger symbols are dwarfs from the higher resolution MARVEL-ous dwarf volumes.  Dwarfs which host black holes span the full range in stellar and halo masses of classical dwarfs, but are notably absent in the ultrafaint mass ranges.  \label{fig:smhm}
}
\end{figure}

An examination of the stellar mass - halo mass relation shows that dwarfs hosting MBHs do not differ from those without (Figure \ref{fig:smhm}). Galaxies which host MBHs are represented by black stars, while grey circles represent the full sample of dwarfs. Point size is indicative of mass-resolution; the lower resolution Justice League dwarfs are smaller points, while the higher resolution MARVEL-ous dwarfs are larger points. Filled stars correspond to isolated or ``field'' dwarfs, and open stars are satellites (as defined in \S \ref{sect:properties}).   Lines are published stellar-to-halo mass relationships from both simulations and abundance matching results, as indicated on the figure. The MBH-hosting dwarfs fill the same region as those without MBHs, within the scatter, indicating that hosting an MBH does not strongly affect the evolution of the stellar properties of dwarfs. Furthermore, hosts span the full dynamic range of stellar and halo masses of classical dwarfs but are absent in mass ranges corresponding to ultra-faint dwarfs.  The environmental trend seen in Figure \ref{fig:occfrac} is also seen here; open stars (satellites) tend to exist above the trend lines, suggesting that they have a higher stellar mass for a given halo mass due to tidal stripping.
\nocite{Moster13,Brook14,Behroozi13,Sawala15}

\begin{figure}
\includegraphics[width=\columnwidth]{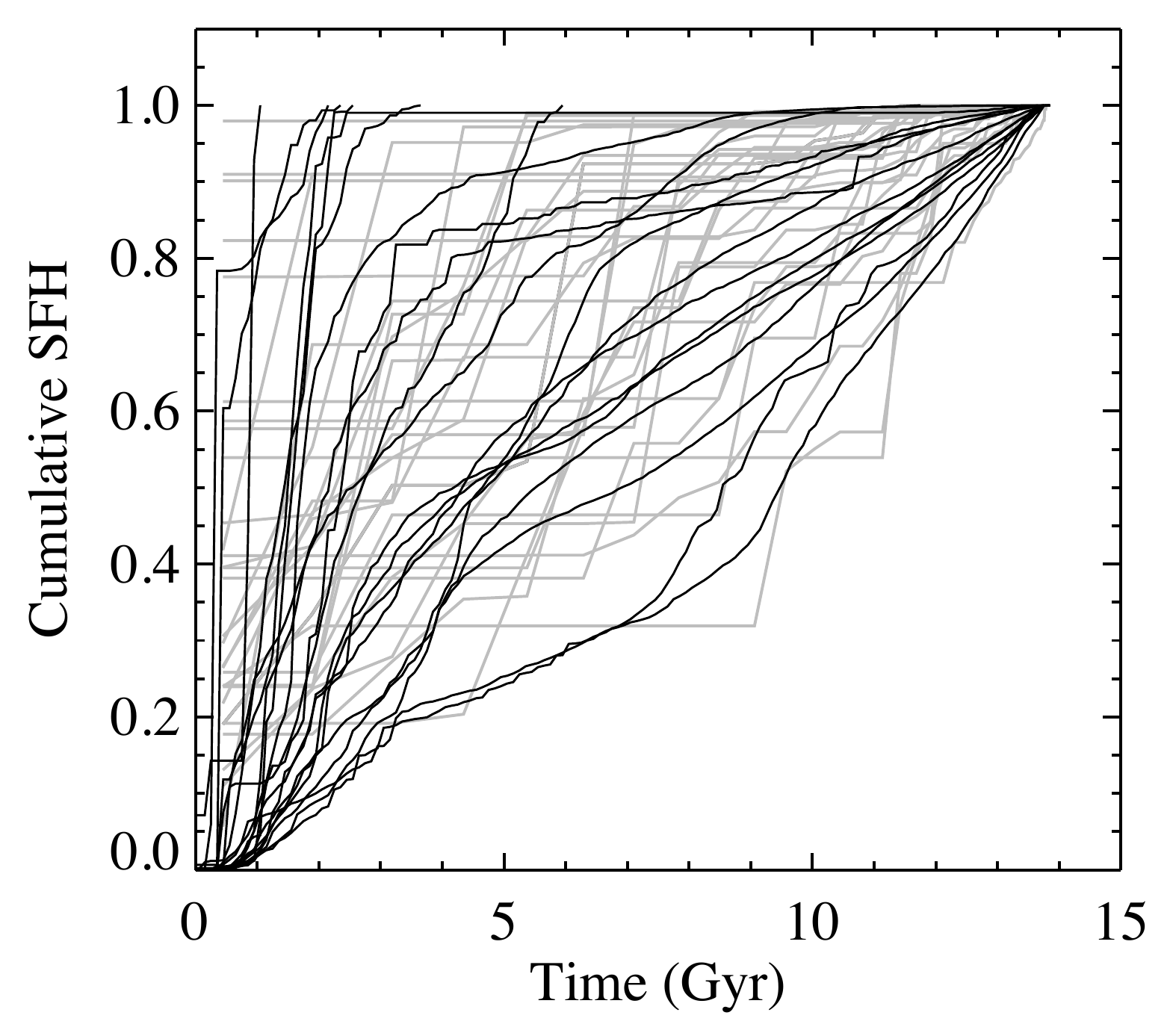}
\caption{ Cumulative star formation histories for dwarf galaxies vs. time.  Black lines are the cumulative stellar mass over time of dwarfs in our simulations.  Grey lines are observed star formation histories of local group dwarfs from \citet{Weisz14}.  The black lines fall within the range consistent with observations. \label{fig:sfhs}
}
\end{figure}

The star formation histories of MBH-hosting dwarfs do not differ from their counterparts without MBHs either.  In Figure \ref{fig:sfhs}, we show the cumulative star formation histories for all simulated dwarf galaxies hosting MBHs with black lines.  The grey lines are the star formation histories of local group dwarf galaxies measured by \citet{Weisz14}.  The black lines from our simulations fall within the broad space outlined by the diverse set of observed histories, confirming that if a dwarf galaxy hosts an MBH with a meager accretion history (based on the low measured luminosities, see \S \ref{sect:detect}), the host properties are not strongly affected.  While one might expect a burst of AGN activity to cause some amount of quenching of star formation, we cannot confirm this phenomenon with our current sample.  

\section{Detecting MBHs in Dwarf Galaxies}\label{sect:detect}

\begin{figure}
\includegraphics[width=\columnwidth]{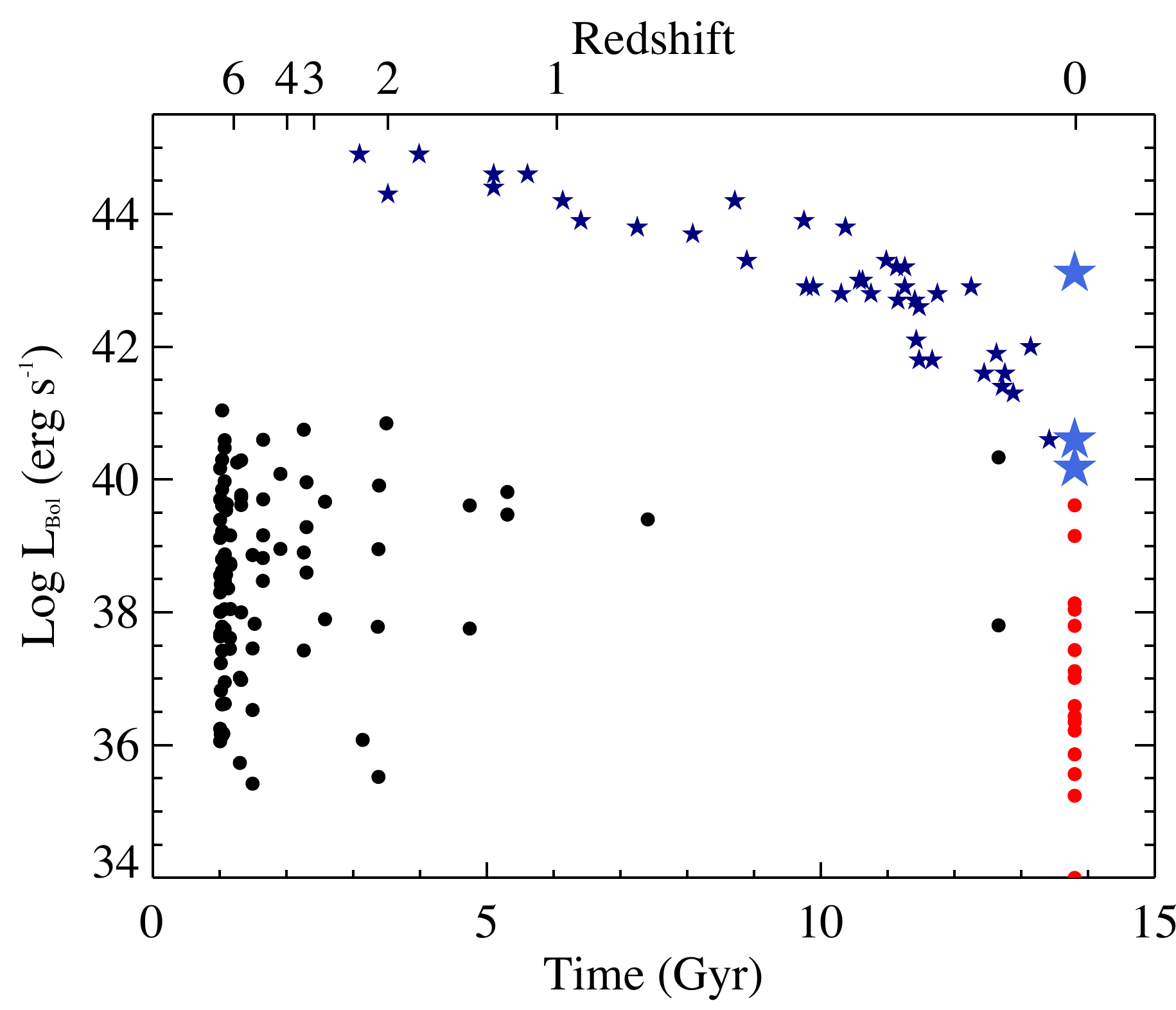}
\caption{ Bolometric luminosities vs time for simulated (circles) and observed (stars) MBHs.  Black circles represent the maximum luminosity that all simulated MBHs have reached at any time between $z = 6$ and the present, smoothed in 10 Myr increments.  
Red circles represent the $z=0$ bolometric luminosities of all simulated MBHs hosted by dwarf galaxies.  Black points outnumber red points due to galaxy and MBH mergers.  Dark blue stars are dwarf-hosted MBH candidates reported in \citet{Mezcua18} from the Chandra COSMOS survey.  Larger light blue stars indicate local MBHs hosted by dwarfs (from the top down, Pox 52, RGG 118, and NGC 4395).  \label{fig:lums}
}
\end{figure}

The electromagnetic signatures of these MBHs due to accretion are quite weak (Figure \ref{fig:lums}).  This weakness is partially a result of their non-central locations, but accretion events are rare even for central MBHs.    The black circles represent the maximum bolometric luminosity of every MBH in every simulation, smoothed in 10 Myr increments, limited to times between $z = 6$ and the present.  
None show luminosities high enough to be characterized as AGN.  The red circles are the bolometric luminosities of every MBH hosted by a dwarf galaxy at $z=0$.  (The black points outnumber the red points due to galaxy and MBH mergers.)  For comparison, the larger light blue stars are local MBHs hosted by dwarfs (from the top down, Pox 52 \citep{Thornton08}, RGG 118 \citep{Baldassare15}, and NGC 4395 \citep{Filippenko93}), which overall exhibit higher luminosities than those in our simulations.  The dark blue stars represent dwarf-hosted MBH candidates from \citet{Mezcua18}, which are all much more luminous than any MBH in our simulations as well.  Overall, the majority of our simulated MBHs are not detectable by a currently existing (or planned future) observatory, while the brightest would be difficult to differentiate from X-ray binaries or other energetic activity.  The observed fraction of dwarfs hosting AGN is less than $\sim 1\%$ \citep{Reines13,Pardo16,Mezcua18}, which is quite low and consistent with our non-detection.  Our results also agree with evidence that low-mass galaxies may host undetected populations of sub-Eddington MBHs, based on X-ray stacking analysis by \citet{Mezcua16}.

We are confident that our model can produce AGN in low-mass galaxies if the conditions are right.  The Romulus simulation, which is a large volume with thousands of low-mass galaxies, hosts several luminous AGN in dwarf galaxies \citep[][Sharma et al in prep.]{Tremmel17}.  Romulus uses the same MBH seeding prescription and physical models.  Our results point to a scenario where the occupation fraction of MBHs in dwarf galaxies is far higher than the AGN fraction, and the conditions for rapid accretion are extremely rare. 
 
While MBHs hosted by dwarf galaxies are unlikely to be observed electromagnetically, their activity will be detectable in the GW regime.  We present here the first prediction for LISA events based on cosmological simulations of low-density galaxy environments.   LISA is especially sensitive to GWs emitted by merging MBHs in the (total) mass range of $10^4 - 10^7$\msun~ up to redshifts of $\sim 20$  \citep{LISA}.  While the precise mass of MBH seeds is unknown, this mass range likely encompasses the majority of MBH seed mergers in the early universe.   

%

\begin{figure}
\includegraphics[width=\columnwidth]{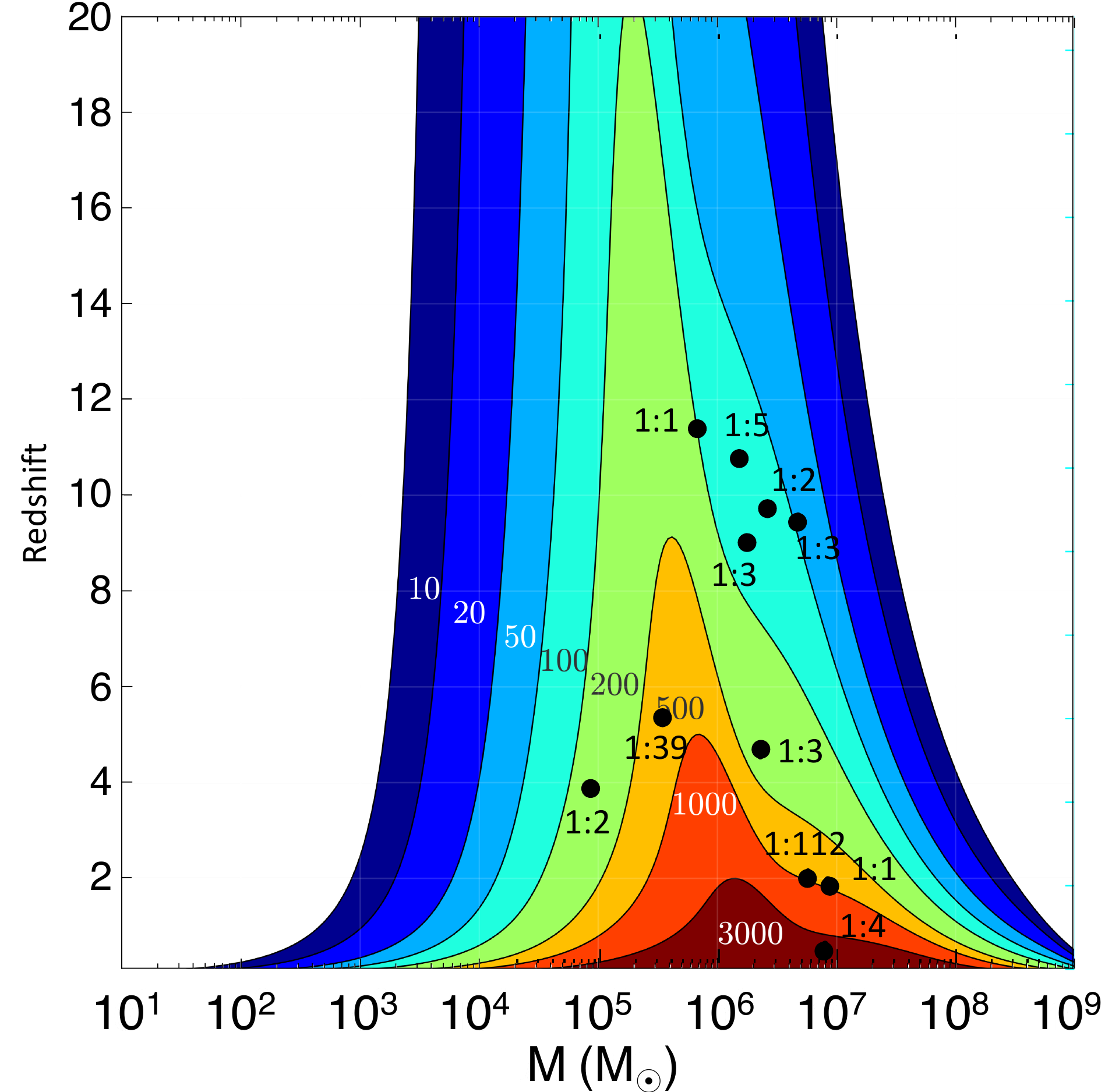}
\caption{ Black points show MBH merger events which take place in dwarf galaxies, as a function of redshift and the combined MBH mass.  Noted near each black point is the mass ratio of each merger.  The rainbow contours and smaller numbers represent the S/N ratio with which LISA will detect such mergers, if they had a characteristic mass ratio of 1:4 (see text).
  \label{fig:waterfall}
}
\end{figure}

In Figure \ref{fig:waterfall} we show the approximate detectability of each MBH merger within a dwarf galaxy with LISA.  The black points are each one MBH-MBH merger which takes place within a dwarf galaxy.  The x-axis shows the combined mass of the two MBHs, while the y-axis is the redshift of the merger event.  Noted near each black point is the mass ratio of each merger.  The coloured contours show the approximate signal-to-noise ratio with which LISA would detect a 1:4 mass ratio merger event, with the black and white numbers quantifying the S/N (background figure courtesy of Martin Hewitson and the LISA Consortium).  The S/N ratio is approximate because it changes as the mass ratio changes.  For larger mass ratios, the S/N increases, and for smaller mass ratios it decreases.  For the majority of the points on the plot, the change is minimal.  This figure shows that MBH mergers in dwarf galaxies occur throughout cosmic time, starting at $z \sim 12$ and continuing to the present.  Most of these mergers occur during epochs when these small, faint galaxies are not observable electromagnetically.  Detecting GWs from these high redshifts will be one of our only clues about low-mass galaxy and structure formation during this epoch.  

The two points with intermediate mass ratios (1:39 at $z \sim 5$ and 1:112 at $z \sim 2$), however, have S/N ratios which are much harder to detect and are not reflected accurately on this figure.  MBH mergers with mass ratios more extreme that 1:10 (but less extreme than 1:1000) are known as ``intermediate mass ratio inspirals'' or IMRIs, and their waveforms are very computationally intensive to calculate, and at the moment are unknown.  Thus, a S/N estimate is not available.  However, as mentioned in Section \ref{sect:Limitations}, some MBHs form with masses much higher than the seed mass in our simulations, due to post-natal overmerging when MBHs form in a clump.  In both of these cases, the larger MBH in the merger is a product of such overmerging.  These two MBHs have masses which are orders of magnitude larger than expected based on MBH-host scaling relations, and are thus unrealistic.  We caution the reader to take these two mass ratios with a grain of salt.

\begin{figure}
\includegraphics[width=\columnwidth]{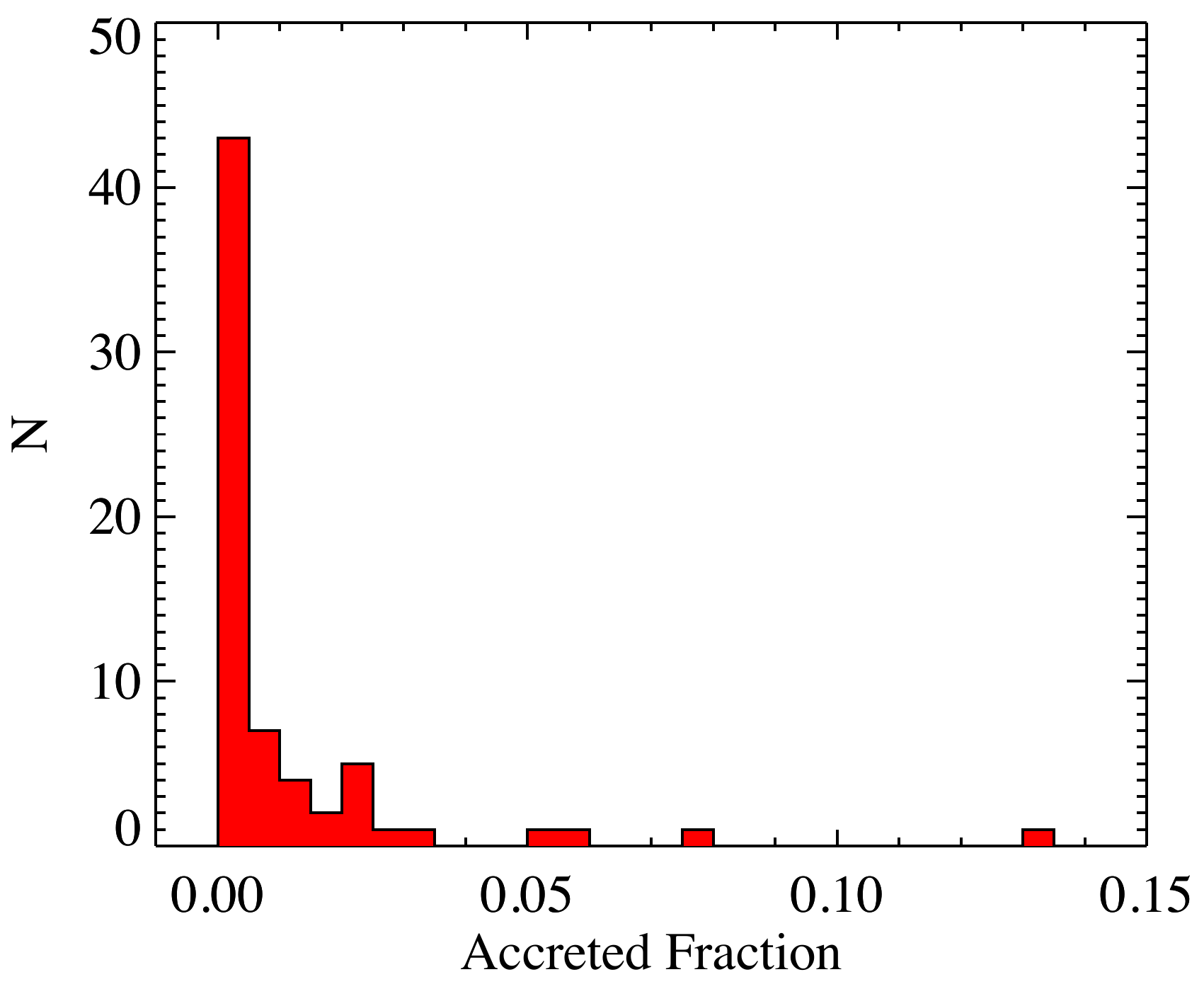}
\caption{ Distribution of the accreted fraction of the total MBH mass ($f_{\rm acc} = M_{\rm acc}/M_{\rm tot}$) for all MBHs in every simulation.  The distribution is heavily weighted toward low values. 
\label{fig:accfrac}
}
\end{figure}

The MBHs in low-mass galaxy environments undergo very little gas accretion and thus their masses are comprised mainly of seed MBHs.  In Figure \ref{fig:accfrac} we show the distribution of the fraction of mass attributed to gas accretion for MBHs which exist at $z=0$ (excluding the central SMBHs of the Milky Way-type galaxies).  The vast majority of these MBHs have accreted very little gas.
  Since they accrete so little, finding a population of such objects can aid in constraining the initial seed mass, leading to hints at the mechanism of seed formation.  Acquiring a sample of  measured MBH masses in dwarf galaxies can provide an upper limit to the seed formation mass \citep{Salcido16}, allowing for a novel way to differentiate seed formation methods besides the detection of gravitational waves from merging seeds \citep{Sesana11,Barausse12}.  Our results motivate the search and discovery of more intermediate mass black holes, in dwarf galaxies and wandering in more massive halos.

This result of minimal black hole growth does not reflect the prevailing thought that SMBHs gain the majority of their gas via gas accretion \citep{Soltan82,Yu02}.  We postulate that such activity occurs in more massive galaxies, where accretable gas is plentiful; however, in lower-mass galaxies, the opportunities for accretion are minimized and thus major accretion events do not take place as frequently.   Our result is consistent with prior works have suggested that MBHs in lower mass galaxies may exhibit ``starved'' histories with little gas accretion  \citep{Dubois15,Bonoli16,Angles-Alcazar17,Bower17,Habouzit17,McAlpine18}.

\section{Summary}

We present the first study of high-resolution cosmological simulations of local dwarf galaxies hosting massive black holes, and discuss the formation, merging, and accretion histories of these MBHs and the effects on their host galaxies. 
 Our results are summarized as follows:

\begin{itemize}
 \item MBH seeds form at high redshifts in low-mass halos.  
 \item About half of MBHs hosted by dwarf galaxies are located in their galaxy centers at $z = 0$ (defined as $< 400$pc), while the remainder are ``wandering'' in the galaxy's outskirts.  
 \item Both the wandering and central MBHs in dwarfs have extremely low accretion luminosities at $z = 0$, rendering them undetectable by modern or even future observatories.  
 \item Dwarf galaxies which host MBHs have similar stellar properties compared to those which do not. 
 \item MBH-MBH mergers in dwarf galaxy environments occur at all redshifts, and their GWs will be in the frequency range detectable by LISA.
 \item MBHs in dwarfs accrete very little gas, thus their masses give clues to the process of seed formation in the early universe.
\end{itemize}

This work is a pivotal first step in studying MBHs in low-mass galaxy environments, but further work must be done.  For example, we neglect the effects of gravitational recoil on MBH-MBH mergers.  The occupation fraction we present in Figure \ref{fig:occfrac} is at best an upper limit, because of the kicks MBHs receive upon merging.  A fraction of these kicks will be enough to eject the MBHs from their hosts \citep[see e.g.][]{Holley-Bockelmann08,Blecha16}.  In future work we will address the effect of gravitational recoil with additional modeling. 

We do not include an estimate of the rates of LISA-observable mergers in this work due to the nature of our study.  The use of the ``zoom-in'' technique results in high resolution but a small sample size.  A statistical estimate is not possible without a full cosmological sample.  In future work we hope to expand our simulation set as well as make connections to the Romulus simulation, which may allow for meaningful GW event rate predictions.

Gravitational waves may be the only way to detect the formation of galaxy and structure formation at the highest redshifts, and will also be instrumental in determining the true formation mechanism(s) of MBH seeds.  Observing local MBHs with electromagnetic radiation is not impossible, as recent studies have shown; however, the fraction of dwarf galaxies hosting AGN is less than 1\% \citep{Reines13,Pardo16,Mezcua18}.  Accreting MBHs in dwarfs are a rare event, and the non-detection of luminous AGN from our simulations is consistent with this low fraction.
Thus LISA will be vital to addressing the existence of intermediate mass black holes in the early and local universe.  In a companion paper, we will examine the specific gravitational wave signatures of individual MBH mergers in more depth.

\section*{Acknowledgements}

This work was supported by the AstroCom NYC project (NSF AST-1153335).  JMB acknowledges support from PSC-CUNY grant 60303-00 48.   AB acknowledges support from HST-AR-14281.  We are grateful to Dan Weisz for supplying local group star formation histories, and Amy Reines and Vivienne Baldassare for helpful comments.  We thank the referee for helpful comments which improved the paper.  Resources supporting this work were provided by the NASA High-End Computing (HEC) Program through the NASA Advanced Supercomputing (NAS) Division at Ames Research Center.


\bibliographystyle{mnras}

\bsp	
\label{lastpage}
\end{document}